\documentstyle[graphicx,prl,aps]{revtex}
   \newcommand{\vecbm}[1]{\mbox{$\boldmath#1$}}

\begin{document}
\paperwidth =15cm
\draft
\title{Second Law of Thermodynamics and Macroscopic Observables\\
within Boltzmann's principle, an attempt} 
\author{D.H.E. Gross} \address{ Hahn-Meitner-Institut
  Berlin, Bereich Theoretische Physik,Glienickerstr.100\\ 14109
  Berlin, Germany and Freie Universit{\"a}t Berlin, Fachbereich
  Physik; \today} 
\maketitle
\begin{abstract} 
  Boltzmann's principle $S=k\ln W$ is generalized to non-equilibrium
  Hamiltonian systems with possibly fractal distributions in phase
  space by the box-counting volume.  The probabilities $P(M)$ of
  macroscopic observables $\hat{M}$ are given by the ratio
  $P(M)=W(M)/W$ of these volumes of the sub-manifold ${\cal{M}}$ of the
  microcanonical ensemble with the constraint $M$ to the one without.
  With this extension of the phase-space integral the Second Law is
  derived without invoking the thermodynamic limit. The
  irreversibility in this approach is due to the replacement of the
  phase space volume of the possibly fractal sub-manifold ${\cal{M}}$
  by the volume of its closure $\overline{{\cal{M}}}$. In contrast to
  conventional coarse graining the box-counting volume is defined by
  the limit of infinite resolution.
\end{abstract}
\section{Introduction}

\noindent Einstein considers Boltzmann's definition of entropy
as e.g. written on his famous epitaph
\begin{equation}
\fbox{\fbox{\vecbm{$S=k$\cdot$lnW$}}}\label{boltzmentr1}\end{equation}
as the fundamental microscopic definition of entropy,  Boltzmann's principle,
\cite{einstein05d}. 
Here $W$ is the number of micro-states at given energy $E$ of the $N$-body 
system in the spatial volume $V$:
\begin{eqnarray}
W(E,N,V)&=& tr[\epsilon_0\delta(E-\hat H_N)]\label{partitsum}\\
tr[\delta(E-\hat H_N)]&=&\int_{\{q\subset V\}}{\frac{1}{N!}
\left(\frac{d^3q\;d^3p}
{(2\pi\hbar)^3}\right)^N\delta(E-\hat H_N)},\label{phasespintegr}
\end{eqnarray} 
$\epsilon_0$ is a suitable energy constant to make $W$ dimensionless,
and $\hat H_N$ is the $N$-particle Hamilton-function.  In what follows, we
remain on the level of classical mechanics. The only reminders of the
underlying quantum mechanics are the measure of the phase space in
units of $2\pi\hbar$ and the factor $1/N!$ which respects the
indistinguishability of the particles (Gibbs paradoxon). In contrast
to Boltzmann \cite{boltzmann1877,boltzmann1884} who used the principle
only for diluted gases and to Schr\"odinger \cite{schroedinger44}, who
thought equation (\ref{boltzmentr1}) is useless otherwise, I take the
principle as {\em the fundamental, generic definition of entropy}. In
a recent book \cite{gross174} cf. also \cite{gross173,gross175} I
demonstrated that this definition of thermo-statistics works well
especially also at higher densities and at phase transitions
without invoking the thermodynamic limit.

After succeeding to deduce all phenomena of phase transitions from
Boltzmann's principle even for ``Small'', i.e. non-extensive many-body
systems, it is challenging to explore how far this ``most
conservative and restrictive way to Thermodynamics''
\cite{bricmont00} is able to describe also (eventually ``Small'')
systems {\em approaching} equilibrium and the Second Law of
Thermodynamics. In the following such an attempt is undertaken.

Thermodynamics describes the development of macroscopic features of
many-body systems without specifying them microscopically to all
details. Before we address the Second Law, we have to clarify what
we mean with the label ``macroscopic observable''.

\section{Measuring a macroscopic observable}
A single point $\{q_i(t),p_i(t)\}_{i=1\cdots N}$ in the $N$-body phase
space corresponds to a detailed microscopic specification of the
system with all degrees of freedom (d.o.f.)  completely fixed at time
$t$ (microscopic determination). Fixing only the total energy $E$ of
an $N$-body system leaves the other $(6N-1)$-degrees of freedom
unspecified. A second system with the same energy is most likely not
in the same microscopic state as the first, it will be at another
point in phase space, the other d.o.f will be different.  I.e.  the
measurement of energy $\hat H$, or any other macroscopic observable
$\hat M$, determines a {\em submanifold} ${\cal E}$ or ${\cal M}$ in
phase space.  All points in phase space which are consistent with the
given value of $E$, i.e. all points in the $(6N-1)$-dimensional
submanifold ${\cal E}$ of the phase space are equally consistent with
this measurement.  ${\cal E}(N,V)$ is the micro-canonical ensemble.
This example tells us that any macroscopic measurement defines a
smaller or greater submanifold of points in phase space. An
additional measurement of any other macroscopic quantity $\hat
B\{q,p\}$ reduces ${\cal{E}}$ further to the cross-section
${\cal{E}}\cap{\cal{B}}$, a subset of points in ${\cal{E}}$ with the
volume:
\begin{equation}
W(B,E,N,V)=\frac{1}{N!}\int{\left(\frac{d^3q\;d^3p}
{(2\pi\hbar)^3}\right)^N\epsilon_0\delta(E-\hat H_N\{q,p\})\;
\delta(B-\hat B\{q,p\})}
\label{integrM}\end{equation}
If $\hat H_N\{q,p\}$ as also $\hat B\{q,p\}$ are continuous
differentiable functions of their arguments, what we assume in the
following, ${\cal{E}}\cap{\cal{B}}$ is closed.

Microcanonical thermo{\em statics} gives the probability $P(B,E,N,V)$
of the $N$-body system, specified only by the prescribed energy $E$,
to be found in the submanifold
${\cal{B}}(E,N,V)\subset{\cal{E}}(N,V)$:
\begin{equation}
P(B,E,N,V)=\frac{W(B,E,N,V)}{W(E,N,V)}=e^{ln(W(B,E,N,V))-S(E,N,V)}\label{prob}
\end{equation}
This is what Krylov seemed to have in mind \cite{krylov79}.

Similarly thermo{\em dynamics} describes the development of some
macroscopic observables $\hat{B}\{q_t,p_t\}$, i.e. the probability
$P(B(t))$ of the system to be at time $t$ in the submanifold
${\cal{B}}(t)$ after it was specified by another macroscopic
observable $\hat{A}\{q_0,p_0\}$ at the earlier time $t_0$. It is
related to the volume of the submanifold
${\cal{M}}(t)={\cal{A}}(t_0)\cap{\cal{B}}(t)\cap{\cal{E}}$:
\begin{equation} W(A,B,E,t)=\frac{1}{N!}\int{\left(\frac{d^3q_t\;d^3p_t}
{(2\pi\hbar)^3}\right)^N\delta(B-\hat B\{q_t,p_t\})\;
\delta(A-\hat A\{q_0,p_0\})\;\epsilon_0\delta(E-\hat H\{q_t,p_t\})},
\label{wab}
\end{equation}
where $\{q_t\{q_0,p_0\},p_t\{q_0,p_0\}\}$ is the set of trajectories
solving the Hamilton-Jacobi equations
\begin{equation}
\dot{q}_i=\frac{\partial\hat H}{\partial p_i},\hspace{1cm}
\dot{p}_i=-\frac{\partial\hat H}{\partial q_i},
\end{equation}
with the initial conditions $\{q(t=t_0)=q_0;\;p(t=t_0)=p_0\}$.

For a very large system with $N\sim 10^{23}$ the probability
$P(B(t))$ is usually sharply peaked as function of $B$. Ordinary
thermodynamics treats systems in the thermodynamic limit $N\to\infty$
and gives only $<\!\!B(t)\!\!>$.  However, here we are interested in
a general application eventually also to small systems i.e. we are
interested in the whole distribution $P(B(t))$ not only in its mean
value $<\!\!B(t)\!\!>$.  Thermodynamics does {\em not} describe the
temporal development of a {\em single} system (single point in the
$6N$-dim phase space).

There is an important property of macroscopic measurements: Whereas
the macroscopic constraint $\hat A\{q_0,p_0\}$ determines (usually) a
compact region ${\cal{A}}(t_0)$ in its arguments this does not need
to be the case for ${\cal{A}}(t)$ given by $\hat
A\{q_0\{q_t,p_t\},p_0\{q_t,p_t\}\}$ as function of $\{q_t,p_t\}$ at
time $t\gg t_0$. It is likely that ${\cal{A}}(t)$ is a {\em fractal}
(e.g.  spaghetti-like) submanifold of $\{q_t,p_t\}$ in ${\cal{E}}$.
I.e. there are series of points $a_n\in {\cal{A}}(t)$ which converge
to a point $a_\infty$ {\em not in} ${\cal{A}}(t)$.  E.g. such points
may have intruded frome outside. Nice examples for this evolution to
fractal distributions in phase space are given e.g.  by
ref.\cite{fox98,gilbert00}. Then no macroscopic (incomplete)
measurement at time $t$ can resolve $a_\infty$ from its immediate
neighbors $a_n$ in phase space with distancees $|a_n-a_\infty|$ less
than any arbitrary small $\delta$. In other words, at the time $t\gg
t_0$ no macroscopic measurement with its incomplete information about
$\{q_t,p_t\}$ can decide whether $q_0\{q_t,p_t\},p_0\{q_t,p_t\}\in
{\cal{A}}(t_0)$ or not.  If necessary, the submanifold determined by
a macroscopic theory like thermodynamics must be the artificially
{\em closed} ${\overline{\cal{M}}(t)}$.  Clearly, this is the origin
of irreversibility. We come back to this in the next section.
\section{Fractal distributions in phase space, Second Law}
Here we will first describe a simple working-scheme (i.e. a sufficient
method) which allows to deduce mathematically the Second Law. Later,
we will show how this method is necessarily implied by the reduced
information obtainable by macroscopic measurements.

Let us examine the following Gedanken experiment: Suppose the
probability to find our system at points $\{q_t,p_t\}_1^N$ in phase
space is uniformly distributed for times $t<t_0$ over the submanifold
${\cal{E}}(N,V_1)$ of the $N$-body phase space at energy $E$ and
spatial volume $V_1$. At time $t_0$ we allow the system to spread over the
larger volume $V_2>V_1$ without changing its energy.  If the system
is {\em dynamically mixing}, i.e.: the majority of trajectories
$\{q_t,p_t\}_1^N$ in phase space starting from points $\{q_0,p_0\}$
with $q_0\subset V_1$ at $t_0$ will now spread over the larger volume
$V_2$.  Of course the Liouvillean measure of the distribution
${\cal{M}}\{q_t,p_t\}$ in phase space at $t>t_0$ will remain the same
($=tr[{\cal{E}}(N,V_1)]$) \cite{goldstein59}:
\begin{eqnarray}
\left.tr[{\cal{M}}\{q_t(q_0,p_0),p_t(q_0,p_0)\}]
\right|_{\{q_0\subset V_1\}}
&=&\int_{\{q_0\{q_t,p_t\}\subset V_1\}}{\frac{1}{N!}\left(\frac{d^3q_t\;d^3p_t}
{(2\pi\hbar)^3}\right)^N\epsilon_0\delta(E-\hat H_N\{q_t,p_t\})}\nonumber\\
&=&\int_{\{q_0\subset V_1\}}{\frac{1}{N!}\left(\frac{d^3q_0\;
d^3p_0}{(2\pi\hbar)^3}\right)^N\epsilon_0\delta(E-\hat H_N\{q_0,p_0\})},\\
\mbox{because of: }\frac{\partial\{q_t,p_t\}}{\partial\{q_0,p_0\}}&=&1.
\end{eqnarray}
But as already argued by Boltzmann the distribution
${\cal{M}}\{q_t,p_t\}$ will be filamented like ink in water and will
approach any point of ${\cal{E}}(N,V_2)$ arbitrarily close.
${\cal{M}}\{q_t,p_t\}$ becomes dense in the new, larger
${\cal{E}}(N,V_2)$ for times sufficiently larger than $t_0$.
The closure $\overline{{\cal{M}}}$ becomes equal to
${\cal{E}}(N,V_2)$.  This is clearly expressed by Lebowitz
\cite{lebowitz99a}.

In order to express this fact mathematically, {\em we have to redefine
  Boltzmann's definition of entropy eq.(\ref{boltzmentr1}) and
  introduce the following fractal ``measure'' for integrals like
  (\ref{phasespintegr}) or (\ref{integrM}):}
\begin{equation}
M(E,N,t\gg t_0)=
\frac{1}{N!}\int_{\{q_0\{q_t,p_t\}\subset V_1\}}{\left(\frac{d^3q_t\;d^3p_t}
{(2\pi\hbar)^3}\right)^N\epsilon_0\delta(E-\hat H_N\{q_t,p_t\})}
\end{equation}
With the transformation:
\begin{eqnarray}
\int{\left(d^3q_t\;d^3p_t\right)^N\cdots}&=&
\int{d\sigma_1\cdots d\sigma_{6N}\cdots}\\
d\sigma_{6N}&:=&\frac{1}{||\nabla\hat H||}
\sum_i{\left(\frac{\partial\hat H}{\partial
q_i}dq_i+\frac{\partial\hat H}{\partial p_i}dp_i\right)}=
 \frac{1}{||\nabla\hat H||}dE\\
||\nabla\hat H||&=&\sqrt{\sum_i{\left(\frac{\partial\hat H}{\partial
q_i}\right)^2+\sum_i{\left(\frac{\partial\hat H}{\partial p_i}\right)^2}}}\\
M(E,N,t\gg t_0)&=&\frac{1}{N!(2\pi\hbar)^{3N}}
\int_{\{q_0\{q_t,p_t\}\subset V_1\}}
{d\sigma_1\cdots d\sigma_{6N-1}
\frac{\epsilon_0}{||\nabla\hat H||}},
\end{eqnarray}
we replace ${\cal{M}}$ by $\overline{\cal{M}}$ and {\em define} now:
\begin{equation}
M(E,N,t\gg t_0)\to<\!G({\cal{E}}(N,V_2)\!>
*\mbox{vol}_{box}[{\cal{M}}(E,N,t\gg t_0)],\label{boxM1}
\end{equation}
where $<\!G({\cal{E}}(N,V_2))\!>$ is the average of
$\frac{\epsilon_0}{N!(2\pi\hbar)^{3N}||\nabla\hat H||}$ over the
(larger) manifold ${\cal{E}}(N,V_2)$, the closure of
${\cal{M}}(E,N,t\gg t_0)$, and $\mbox{vol}_{box}[{\cal{M}}(E,N,t\gg
t_0)]$ is the box-counting volume of ${\cal{M}}(E,N,t\gg t_0)$.

To obtain $\mbox{vol}_{box}[{\cal{M}}(E,N,t\gg t_0)]$ we cover the
$d$-dim. submanifold ${\cal{M}}(t)$, here with $d=(6N-1)$,  of the
phase space by a grid with spacing $\delta$ and count the number
$N_\delta\propto
\delta^{-d}$ of boxes, of size $\delta^{6N}$, which contain points of
${\cal {M}}$.  Then we determine
\begin{eqnarray}
\mbox{vol}_{box}[{\cal{M}}(E,N,t\gg t_0)]&:=&\underbar{$\lim$}_{\delta\to 0}
\delta^d N_\delta[{\cal{M}}(E,N,t\gg t_0)]\label{boxvol}\\
\lefteqn{\mbox{with }\underbar{$\lim *$}=\inf[\lim *]\mbox{ or
symbolically:}} \nonumber\\
{\cal{M}}(E,N,t\gg t_0)&=:&
\displaystyle{B_d\hspace{-0.5 cm}\int}_{\{q_0\{q_t,p_t\}\subset V_1\}}
{\frac{1}{N!}\left(\frac{d^3q_t\;
d^3p_t}{(2\pi\hbar)^3}\right)^N\epsilon_0\delta(E-\hat
H_N)}\label{boxM}\\ 
&\to&\frac{1}{N!}\int_{\{q_t\subset V_2\}}
{\left(\frac{d^3q_t\;d^3p_t}
{(2\pi\hbar)^3}\right)^N\epsilon_0\delta(E-\hat H_N\{q_t,p_t\})}
\nonumber\\
&=&W(E,N,V_2) \ge W(E,N,V_1),
\end{eqnarray}
where $\displaystyle{B_d\hspace{-0.5 cm}\int}$ means that this
integral should be evaluated via the box-counting volume
(\ref{boxvol}) here with $d=6N-1$.

With this extension of eq.(\ref{phasespintegr}) Boltzmann's entropy
(\ref{boltzmentr1}) is at time $t\gg t_0$ equal to the logarithm of the 
{\em larger} phase space $W(E,N,V_2)$. This is the Second Law of
Thermodynamics.

Of course still at $t_0$ 
$\overline{{\cal{M}}(t_0)}={\cal{M}}(t_0)={\cal{E}}(N,V_1)$:
\begin{eqnarray}
M(E,N,t_0)
&=:&\displaystyle{B_d\hspace{-0.5 cm}\int}_{\{q_0\subset V_1\}}
{\frac{1}{N!}\left(\frac{d^3q_0\;
d^3p_0}{(2\pi\hbar)^3}\right)^N\epsilon_0\delta(E-\hat H_N)}\\
&\equiv&\int_{\{q_0\subset V_1\}}
{\frac{1}{N!}\left(\frac{d^3q_0\;
d^3p_0}{(2\pi\hbar)^3}\right)^N\epsilon_0\delta(E-\hat H_N)}\nonumber\\
&=&W(E,N,V_1).
\end{eqnarray} 

The box-counting volume is analogous to the standard method to
determine the fractal dimension of a set of points \cite{falconer90}
by the box-counting dimension which is itself closely related to the
Kolmogorov entropy \cite{falconer90,crc99}:
\begin{equation}
\dim_{box}[{\cal{M}}(E,N,t\gg t_0)]:=\underbar{$\lim$}_{\delta\to 0}
\frac{\ln{N_\delta[{\cal{M}}(E,N,t\gg t_0)]}}{-\ln{\delta}}\\
\end{equation}

Like the box-counting dimension, $\mbox{vol}_{box}$ has the peculiarity
that it is equal to the volume of the smallest {\em closed} covering
set. E.g.: The box-counting volume of the set of rational numbers
$\{{\bf Q}\}$ between $0$ and $1$, is $\mbox{vol}_{box}\{{\bf Q}\}=1$,
and thus equal to the measure of the {\em real} numbers , c.f.
Falconer \cite{falconer90} section 3.1. This is the reason why
$\mbox{vol}_{box}$ is not a measure in its mathematical definition
because then we should have
\begin{equation}
\mbox{vol}_{box}\left[\sum_{i\subset\{\bf  Q\}}({\cal{M}}_i)\right]=
\sum_{i\subset\{\bf  Q\}}\mbox{vol}_{box}[{\cal{M}}_i]=0,
\end{equation}
therefore the quotation marks for the box-counting ``measure''.

Coming back to the the end of the previous section, the volume
$W(A,B,\cdots,t)$ of the relevant ensemble, the {\em closure}
$\overline{{\cal{M}}(t)}$ must be ``measured'' by something like the
box-counting ``measure'' (\ref{boxvol},\ref{boxM}) with the
box-counting integral $\displaystyle{ B_d\hspace{-0.5 cm}\int}$, which
must replace the integral in eq.(\ref{wab}).
\section{Conclusion}
Macroscopic measurements $\hat{M}$ determine only a very few of all
$6N$ d.o.f.  Any macroscopic theory like thermodynamics deals with
the {\em volumes} of the corresponding submanifolds ${\cal{M}}$ in
the $6N$-dim. phase space not with single points. This fact becomes
especially clear for the microcanonical ensemble of a finite system.
Because of this necessarily coarsed information macroscopic
measurements, and with it also macroscopic theories are unable to
distinguish fractal sets ${\cal{M}}$ from their closures
$\overline{\cal{M}}$. Therefore, the proper manifolds determined by a
macroscopic theory like thermodynamics are the closed
$\overline{\cal{M}}$. However, an initially closed subset of points
at time $t_0$ does not necessarily evolve again into a closed subset
at $t>t_0$. I.e. the closure operation does not commute with the
dynamics of a set of points in phase space, and the macroscopic
dynamics becomes irreversible.

The use of the box-counting volume $\displaystyle{B_d\hspace{-0.5
    cm}\int}$ in Boltzmann's principle (eq.\ref{boltzmentr1} together
with eq.\ref{phasespintegr}) allows to derive the Second Law without
invoking the thermodynamic limit $N\to\infty$. We must only demand
that $N\gg$ than the small number of explicit macroscopic d.o.f.
($\sim 3$), the control parameters.  This is in contrast to Lebowitz
\cite{lebowitz99a}. Also no finite coarse graining is needed.  The
resolution $\delta$ can be chosen arbitrarily small. The prize to be
paid is that $\displaystyle{ B_d\hspace{-0.5 cm}\int}$~ is not a
measure, see above. Evidently, due to this replacement of ${\cal{M}}$
by $\overline{{\cal{M}}}$ or the integral over phase space,
eq.(\ref{phasespintegr}), by its box-counting variant,
eq.(\ref{boxM}), the irreversibility and the increase of entropy comes
about in our formalism.\\
Thanks to A.Ecker for mathematical advices.

\end{document}